\documentclass{sf2a-conf2016}
\usepackage{graphicx,multirow,supertabular,placeins}
\usepackage{hyperref}
\usepackage[]{natbib}  
\usepackage{epstopdf}

\def\BibTeX{{\rm B\kern-.05em{\sc i\kern-.025em b}\kern-.08em
    T\kern-.1667em\lower.7ex\hbox{E}\kern-.125emX}}
\bibpunct{(}{)}{;}{a}{}{,}  

\begin{document}

\TitreGlobal{SF2A 2016}


\title{The WEAVE-LOFAR Survey}

\runningtitle{The WEAVE-LOFAR Survey}

\author{D.J.B.~Smith}\address{Centre for Astrophysics Research, School of Physics Astronomy and Mathematics, University of Hertfordshire, College Lane, Hatfield, Hertfordshire, AL10 9AB, United Kingdom}

\author{P.N.~Best}\address{Institute for Astronomy, Royal Observatory, Blackford Hill, Edinburgh, EH9 3HJ, United Kingdom}


\author{K.J.~Duncan}\address{Leiden Observatory, Universiteit Leiden, Huygens Laboratory\slash J.H. Oort Building, Niels Bohrweg 2, NL-2333 CA Leiden, The Netherlands} 

\author{N.A.~Hatch}\address{The Centre for Astronomy \&\ Particle Theory, School of Physics \&\ Astronomy, Nottingham University, University Park, Nottingham, NG7 2RD, United Kingdom} 

\author{M.J.~Jarvis}\address{Oxford Astrophysics, Denys Wilkinson Building, Keble Road, Oxford, OX1 3RH, United Kingdom} 

\author{H.J.A.~R\"ottgering$^3$}

\author{C.J.~Simpson}\address{Gemini Observatory, Northern Operations Center, 670 North A`oh\={o}k\={u} Place, Hilo, HI 96720-2700, USA}
\author{J.P. Stott$^5$}

\author{R.K.~Cochrane$^2$}
\author{K.E.~Coppin$^1$}
\author{H.~Dannerbauer}\address{Instituto de Astrofísica de Canarias (IAC), E-38205 La Laguna, Tenerife, Spain \&\
Universidad de La Laguna, Dpto. Astrofísica, E-38206 La Laguna, Tenerife, Spain}
\author{T.A.~Davis}\address{School of Physics \&\ Astronomy, Cardiff University, Queens Buildings, The Parade, Cardiff, CF24 3AA, UK}
\author{J.E.~Geach$^1$}
\author{C.L.~Hale$^5$}
\author{M.J.~Hardcastle$^1$}
\author{P.W.~Hatfield$^5$}
\author{R.C.W.~Houghton$^5$}
\author{N.~Maddox}\address{ASTRON, the Netherlands Institute for Radio Astronomy, Postbus 2, 7990 AA, Dwingeloo, The Netherlands}
\author{S.L.~McGee}\address{School of Physics and Astronomy, University of Birmingham, Edgbaston, Birmingham, B15 2TT, UK}
\author{L.~Morabito$^5$}
\author{D.~Nisbet$^2$}
\author{M.~Pandey-Pommier}\address{Univ Lyon, Univ Lyon1, Ens de Lyon, CNRS, Centre de Recherche Astrophysique de Lyon UMR5574, 9 av Charles André, F- 69230, Saint-Genis-Laval, France}
\author{I.~Prandoni}\address{INAF-IRA, Via P. Gobetti 101, 40129 Bologna, Italy}
\author{A.~Saxena$^3$}
\author{T.W.~Shimwell$^3$}
\author{M.~Tarr}\address{ICG, Dennis Sciama Building, Burnaby Road, Portsmouth, PO1 3FX, United Kingdom}
\author{I.~van~Bemmel}\address{Joint Institute for VLBI ERIC, PO Box 2, 7990 AA Dwingeloo, The Netherlands}
\author{A.~Verma$^5$}
\author{G.J.~White}\address{Department of Physics and Astronomy, The Open University, Walton Hall, Milton Keynes, MK7 6AA, UK \&\ RAL Space, STFC Rutherford Appleton Laboratory, Chilton, Didcot, Oxfordshire, OX11 0QX, UK}
\author{W.L.~Williams$^1$}
\author{the~WEAVE~Consortium}.

\setcounter{page}{1}


\maketitle


\begin{abstract}
In these proceedings we highlight the primary scientific goals and design of the WEAVE-LOFAR survey, which will use the new WEAVE spectrograph on the 4.2m William Herschel Telescope to provide the primary source of spectroscopic information for the LOFAR Surveys Key Science Project. Beginning in 2018, WEAVE-LOFAR will generate more than $10^6$ R=5000 365-960nm spectra of low-frequency selected radio sources, across three tiers designed to efficiently sample the redshift-luminosity plane, and produce a data set of enormous legacy value. The radio frequency selection, combined with the high multiplex and throughput of the WEAVE spectrograph, make obtaining redshifts in this way very efficient, and we expect that the redshift success rate will approach 100 per cent at $z < 1$. This unprecedented spectroscopic sample -- which will be complemented by an integral field component -- will be transformational in key areas, including studying the star formation history of the Universe, the role of accretion and AGN-driven feedback, properties of the epoch of reionisation, cosmology, cluster haloes and relics, as well as the nature of radio galaxies and protoclusters. Each topic will be addressed in unprecedented detail, and with the most reliable source classifications and redshift information in existence.
\end{abstract}



\section{Introduction \&\ Motivation}

The International LOFAR Telescope offers a transformational increase in survey speed compared to existing radio telescopes, as well as opening up one of the few poorly explored regions of the electromagnetic spectrum. An important driver for LOFAR, since its inception, has been to carry out a series of surveys of the low-frequency radio sky to advance, in particular, our understanding of the formation and evolution of galaxies, clusters, and active galactic nuclei (AGN).
 
The LOFAR Surveys Key Science Project (KSP) is described in \citet{rottgering11}, but to summarize, LOFAR is carrying out surveys in a wedding-cake strategy, with three tiers of observations to be completed over the next $\sim5$ years. Tier-1 is the widest tier, and includes low-band (LBA; between 10-80 MHz) and high-band (HBA; between 120-240 MHz) observations across the whole $2\pi$ steradians of the northern sky to a depth approximately 10 times that of the FIRST survey \citep{becker95}, or $\sim 600$ times deeper than the VLA Low-frequency Sky Survey \citep{cohen07}. Deeper Tier-2 and Tier-3 observations are proposed over smaller areas (Tier-3 will cover $\sim 100$\,deg$^2$ to a depth greater than that of the deepest current radio imaging), focusing on fields with the highest quality multi-wavelength datasets available across a broad range of the electromagnetic spectrum. The effectiveness of new direction-dependent calibration techniques (necessary to account for the large confusing influence of the ionosphere on low-frequencies) has been demonstrated by several publications \citep[e.g.][]{hardcastle16,shimwell16,vanweeren16,williams16}, while LOFAR's ability to rapidly survey large areas has also been demonstrated (Shimwell et al. {\it submitted}).

\begin{figure}[t]
 \centering
 \includegraphics[width=0.8\textwidth,clip]{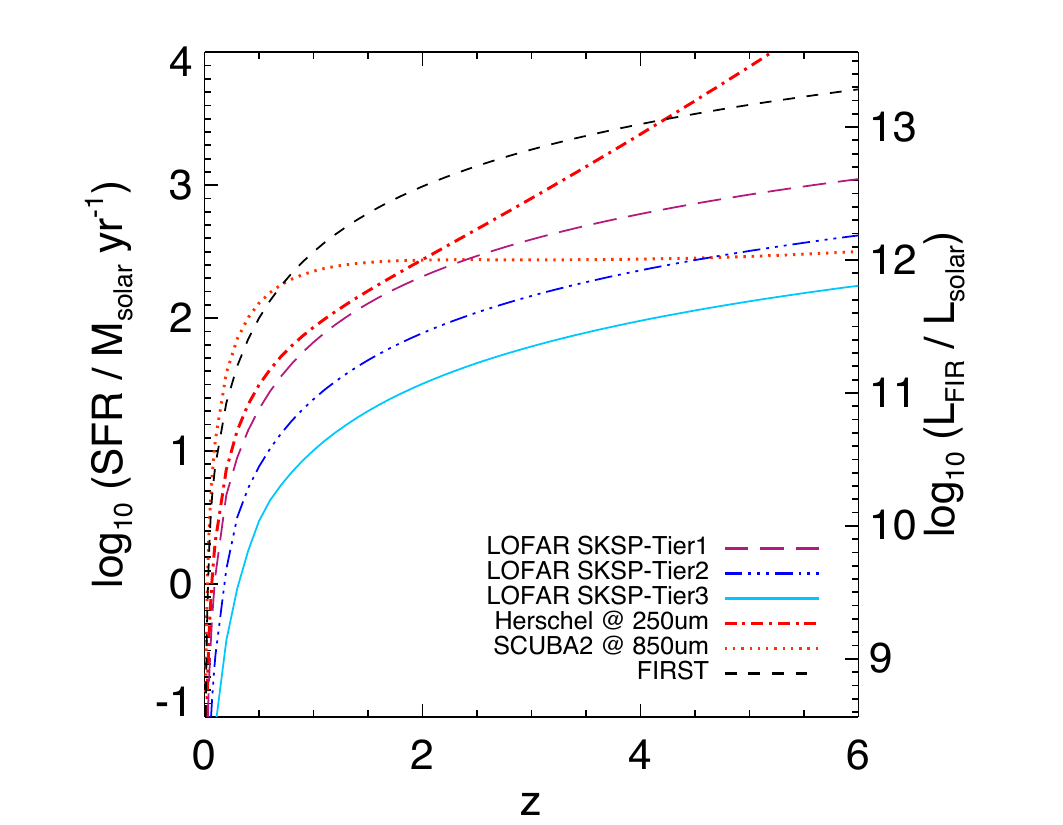}      
  \caption{A comparison between the star formation rate sensitivity of the $5\sigma$ limits of each tier of the LOFAR Surveys KSP (purple long-dashed, blue dot-dot-dot-dashed and light-blue solid lines for Tiers 1, 2 \&\ 3, respectively), and selected other surveys, including FIRST \citep[][shown as the dashed black line]{becker95}, and five times the confusion noise limit of {\it Herschel Space Observatory} at 250\,$\mu$m \citep[taken from][red dot-dashed line, assuming a canonical isothermal template with $T=35K$ and $\beta = 1.5$]{rigby11}, and from SCUBA-2 at 850\,$\mu$m \citep[dotted red line, assuming the same template SED and the confusion noise estimate taken from][]{geach16}. We assume an universal radio spectral index of $\alpha = 0.71$ from \citet{mauch13}, adopting the convention that $S_\nu \propto \nu^{-\alpha}$. }
  \label{fig:dsmith_sfr_sens}
\end{figure}

Whilst the high sensitivity of low-frequency radio surveys to AGN is well known, figure \ref{fig:dsmith_sfr_sens} shows that the LOFAR Surveys KSP data are also supremely sensitive to star-formation, assuming standard relationships between radio flux density and star formation rate (SFR) from \citet{bell03}. LOFAR is able to detect star-forming systems that are beyond the reach of even confusion limited observations with the {\it Herschel Space Observatory} \citep{pilbratt10} and SCUBA-2 \citep[despite the negative $K$-correction at sub-millimetre wavelengths]{holland13}.

The William Herschel Telescope Enhanced Area Velocity Explorer instrument \citep[WEAVE;][]{dalton12,dalton14} is a next-generation spectroscopy facility, which has been designed with follow-up of LOFAR targets as one of the primary goals. WEAVE is a multi-object (MOS) and multi-integral field unit (IFU) fibre-fed spectrograph, which allows 1,000 fibres to be positioned robotically over a field of view 2\,degrees in diameter. In ``low resolution" mode, WEAVE produces spectra at $R=5,000$ over a contiguous wavelength range between 365-960\,nm in a single exposure, ideal for the efficient detection, redshifting and classification of radio sources. 
 
WEAVE-LOFAR is therefore tasked with being the primary source of spectroscopic information for the LOFAR surveys. Spectra are required for much more than simply estimating redshifts; it is only using these data that we are able to robustly distinguish between star forming galaxies (SFGs) and AGN, and between accretion modes in those AGN themselves. Spectra also permit us to measure velocity dispersions, estimate metallicities and derive virial black hole mass estimates; many thousands of WEAVE-LOFAR spectra thus offer an unique insight into the relationship between star formation and accretion over a huge swathe of cosmic history. In addition, since we target sources selected at radio frequencies (which are unaffected by dust obscuration), WEAVE-LOFAR will be much less biased against obscured sources than samples selected at optical/near-infrared wavelengths, providing a representative view of the galaxy and AGN populations. Furthermore, the unique capability offered by the suite of WEAVE integral field units (IFUs) enables us to investigate the onset of cluster formation in the early Universe (putting proto-clusters into cosmological context for the first time), allows us to probe the gas supply of massive high-redshift galaxies, and provides the opportunity to measure the impact of AGN-driven feedback on galaxy evolution.
 
 
 
Some of the key areas that WEAVE-LOFAR will address include:
 
\begin{itemize}
\item the star-formation history of the Universe,
\item accretion and AGN-driven feedback,
\item the epoch of reionisation,
\item cosmology,
\item cluster haloes and relics,
\item radio galaxies and protoclusters.
\end{itemize}

The WEAVE-LOFAR survey will be described in a future publication (Smith et al. in prep), but in section \ref{sec:dsmith_design} we discuss the survey design and expected redshift success rate, while in section \ref{sec:dsmith_cases} we discuss some key areas of the science case in more detail. We conclude in section \ref{sec:dsmith_concs}.

\section{Survey Design}
\label{sec:dsmith_design}

\subsection{Tier strategy and choice of fields} 

WEAVE-LOFAR will adopt a tiered strategy, designed to efficiently sample the redshift-luminosity plane. In this way, the survey will obtain statistical samples of both the rarest bright sources (e.g. radio galaxies in the epoch of reionisation) and the star-forming galaxies that dominate the source counts at the faintest flux densities. WEAVE-LOFAR will have three tiers named Deep, Mid and Wide, which mimic the enumerated tiers of the LOFAR Surveys KSP, though targeting only a subset of the sources in each. The targeting strategy will be: 

\begin{itemize}
	\item Deep: $S_{\mathrm{150 MHz}} > 100\,\mu$Jy, 
	\item Mid: $S_{\mathrm{150 MHz}} > 1$\,mJy,
	\item Wide: $S_{\mathrm{150 MHz}} > 10$\,mJy.
\end{itemize}

\begin{figure}[t]
 \centering
 \includegraphics[height=0.5\textwidth, trim=0cm 0cm 3.2cm 0cm, clip=false]{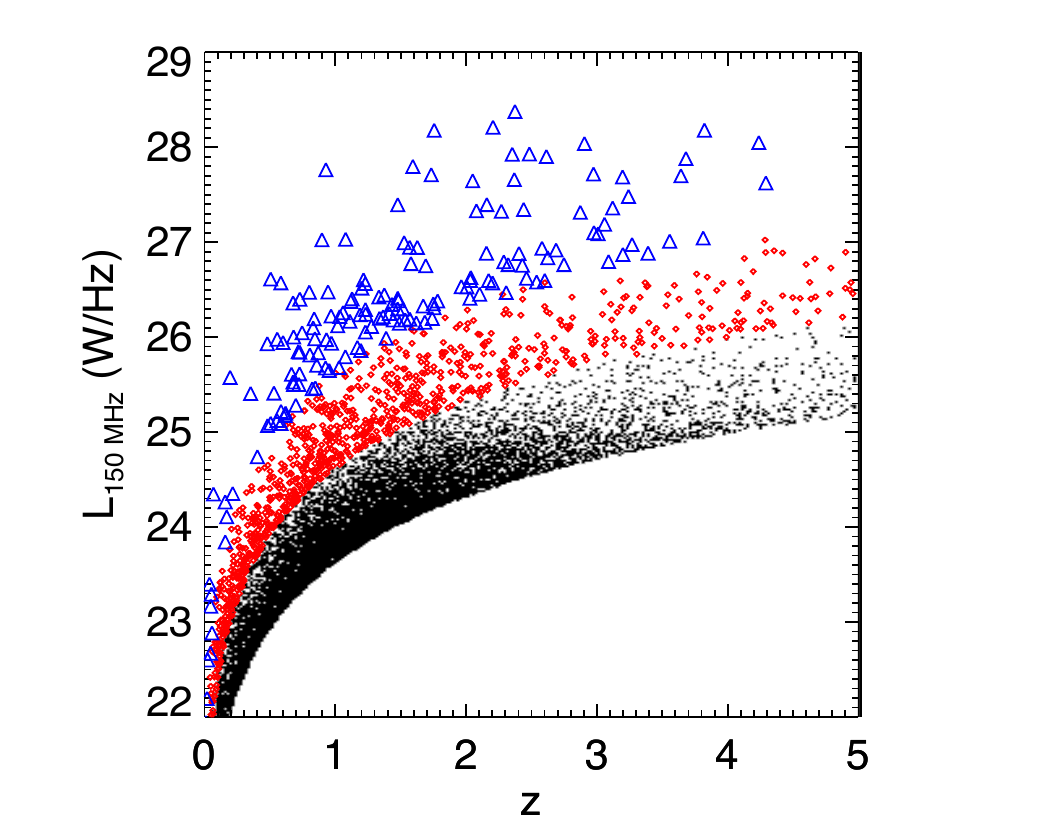}%
 \includegraphics[height=0.5\textwidth, trim=3.2cm 0cm 0cm 0cm, clip=false]{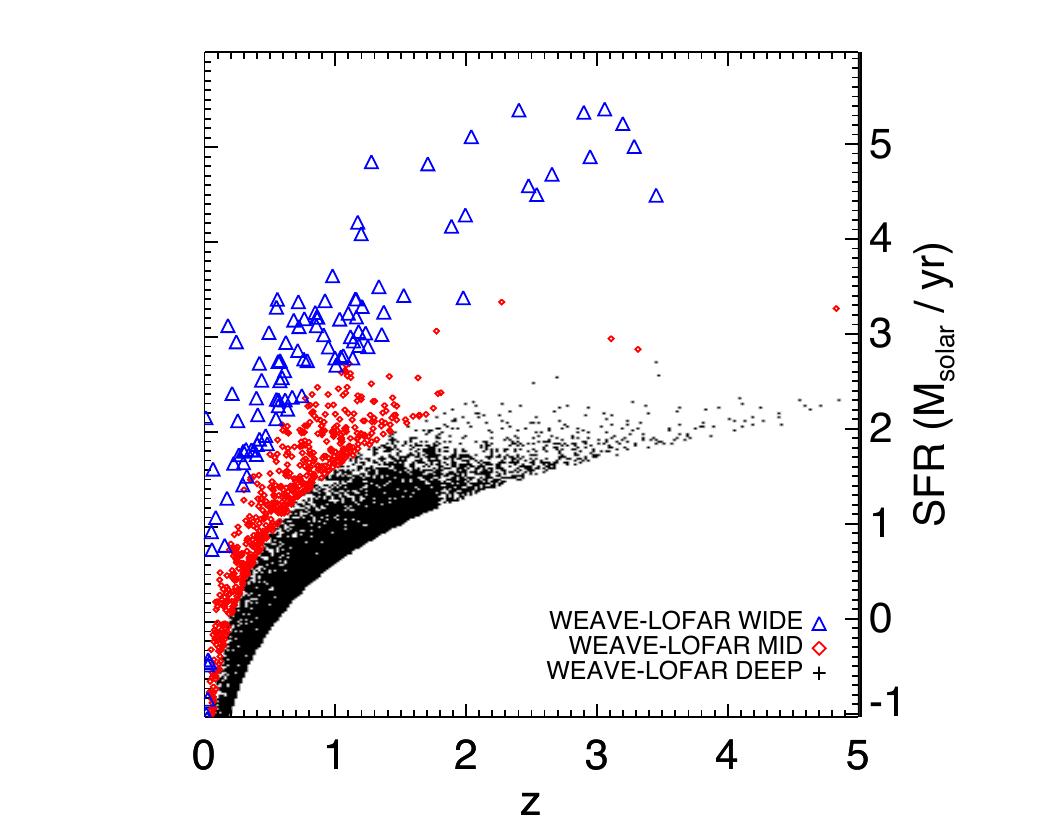}      
  \caption{{\bf Left:} The luminosity redshift plane as sampled by fields drawn from each of the three tiers of the WEAVE-LOFAR survey. Targets in the Wide tier are shown as blue triangles, while those in the mid tier are shown as red diamonds. The faintest, most numerous sources in the Deep tier are shown as black points. The right-hand axis has been scaled to reflect the star-formation rates associated with the values on the left-hand y-axes, though clearly the more luminous sources' radio emission is not powered by star formation, but by AGN. {\bf Right:} As for the left panel, but only those sources for which we expect -- based on our simulations -- to obtain successful redshifts are shown. The expected redshift success rate in each tier is 75.6, 60.8 and 67.1 per cent in the Deep, Mid and Wide tiers, respectively.
   }
  \label{dsmith:lumz_fig}
\end{figure}

\begin{figure}[t]
 \centering
 \includegraphics[width=0.8\textwidth,clip]{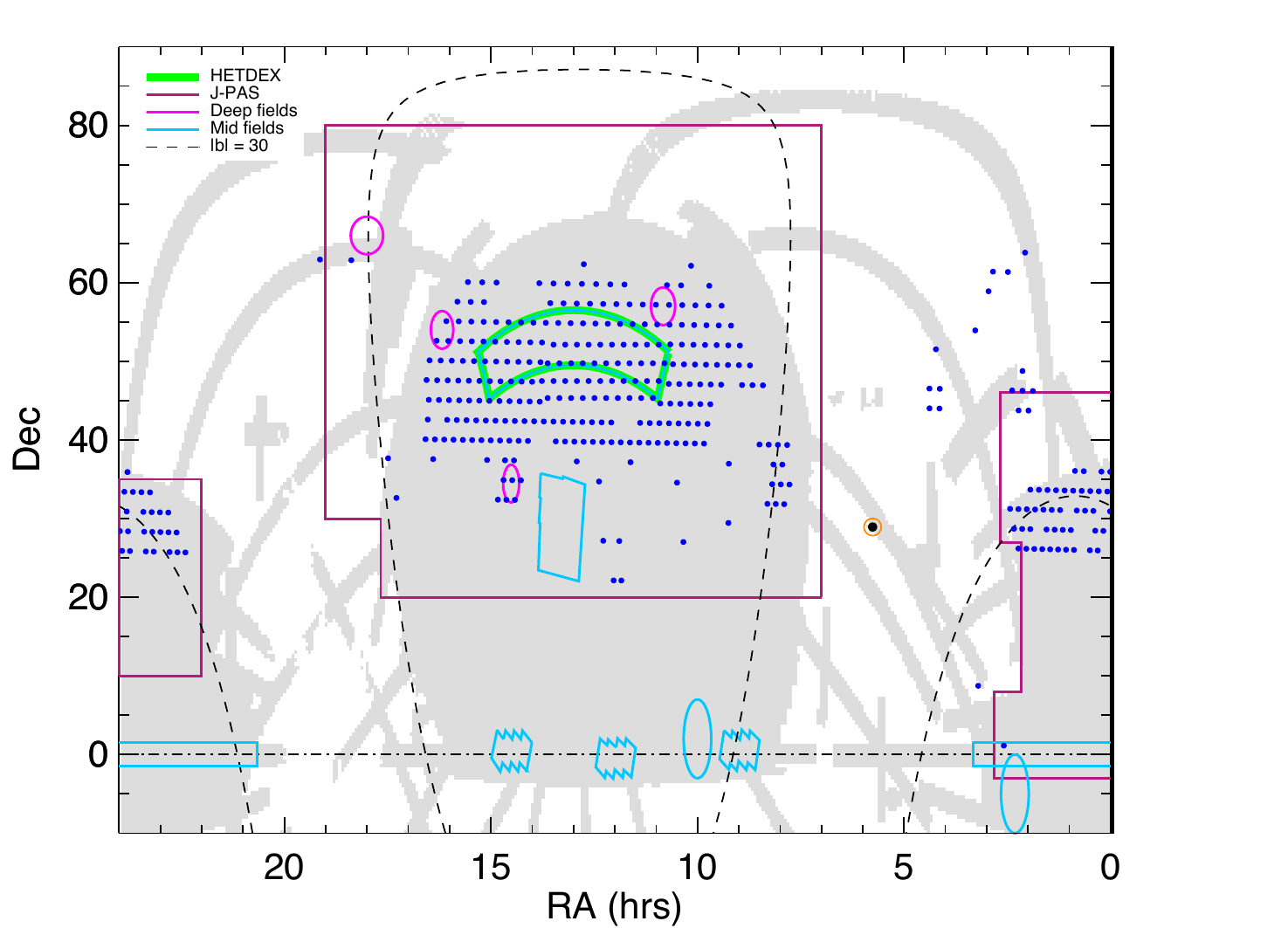}      
  \caption{The primary fields for the WEAVE-LOFAR survey detailed in table \ref{table:weave_fields}, overlaid on the outline of the SDSS imaging (grey). The Mid-tier fields are outlined in light blue, alongside the Deep fields (pink circles). The primary Mid-tier field, the HETDEX northern field \citep{hill08}, is outlined in green, while the outline of the J-PAS survey \citep{benitez14} is shown in purple. The fields observed as part of the LOFAR Surveys KSP up to the end of observing cycle 6 are shown as the blue circles. The black circle outlined in orange shows the location of the galactic anti-centre, while the the black dashed lines correspond to galactic latitudes $|b| = 30$.}
  \label{fig:dsmith_fields}
\end{figure}

\noindent Together, the three tiers fill the redshift-luminosity plane as shown in figure \ref{dsmith:lumz_fig}, which are based on realisations of a single WEAVE field of view for each tier of our survey in the SKA simulated skies from \citet{wilman08}. The source density above the flux limit in each tier is 15,000\slash 1,100\slash 160 per $\pi$\,deg$^2$ WEAVE field of view for the Deep\slash Mid\slash Wide tiers respectively. In the Deep and Mid tiers, this is high enough to enable us to define our own dedicated pointings (since the number of available targets exceeds the number of fibres in the MOS spectrograph). We have chosen those fields visible by both WEAVE and LOFAR that have the best available multi-wavelength ancillary data; these are detailed in table \ref{table:weave_fields}, and plotted in figure \ref{fig:dsmith_fields} (light-blue regions represent the Mid tier fields, while the deep fields are shown as pink circles).  

In the Wide tier, the low source density of the brightest and rarest sources (including samples of high redshift powerful AGN, as well as the ultra-rare radio galaxies within the Epoch of Reionisation, the EoR) means that we must share fibres with other WEAVE surveys to efficiently use the instrument. Fortunately the WEAVE Galactic Archaeology (GA; Hill et al. {\it in prep}) survey and the WEAVE QSO survey (Pieri et al. {\it in prep}) also require wide areal coverage, and we intend to follow this strategy to cover the widest possible area (up to 10,000\,deg$^2$ over the initial five years of survey operations). The LOFAR Surveys KSP observations to date  are shown as the blue circles in figure \ref{fig:dsmith_fields}), and the necessity of observing fields in common with the GA and QSO teams has been a key ingredient in identifying the regions to observe in recent LOFAR cycles. 

\begin{table}
\caption{Field names and approximate centres (in the standard format, HHMM+DD) for the WEAVE-LOFAR Deep and Mid tiers. The Wide tier will be co-located with the WEAVE-GA and WEAVE-QSO surveys, spread over regions in the galactic halo, primarily at $|b| > 30$.}
\vspace{0.3cm}
\centering
\begin{tabular}{llcll}
{\bf Mid Tier:} & & \hspace{2cm} & {\bf Deep Tier:} & \\
\cline{1-2} \cline{4-5}
Name & Pointing Centre & & Name & Pointing Centre \\
\cline{1-2} \cline{4-5}
HETDEX & 1200+55 & & Bo\"otes & 1430+34 \\
{\it H}-ATLAS NGP & 1300+29 & & Lockman Hole & 1050+57 \\
{\it H}-ATLAS GAMA & 0900/1200/1430+00 & & ELAIS-N1 & 1610+54 \\
SDSS Stripe 82 & 0000+00 & & NEP & 1800+66 \\
 \cline{4-5}
 XMM-LSS & 0219-05 & & Total Area: & $\sim$100\,deg$^2$ \\
COSMOS & 1000+02 & & & \\
\cline{1-2}
Total Area: & $\sim$1,250\,deg$^2$
\end{tabular}
\label{table:weave_fields}
\end{table}

\subsection{Expected redshift success rates} 

\begin{figure}[t]
 \centering
 \includegraphics[height=0.27\textwidth, trim=0cm 0cm 0.52cm 0cm, clip=true]{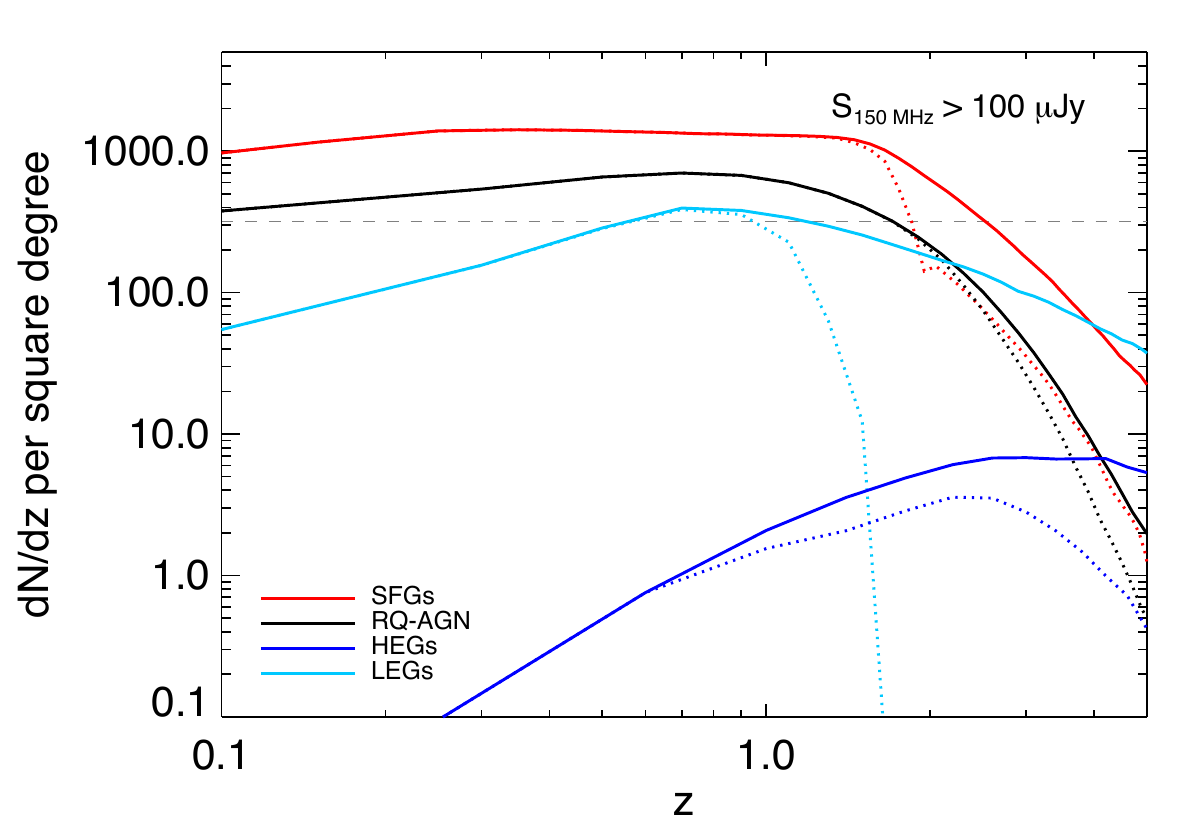}%
 \includegraphics[height=0.27\textwidth, trim=3.8cm 0cm 0.52cm 0cm, clip=false]{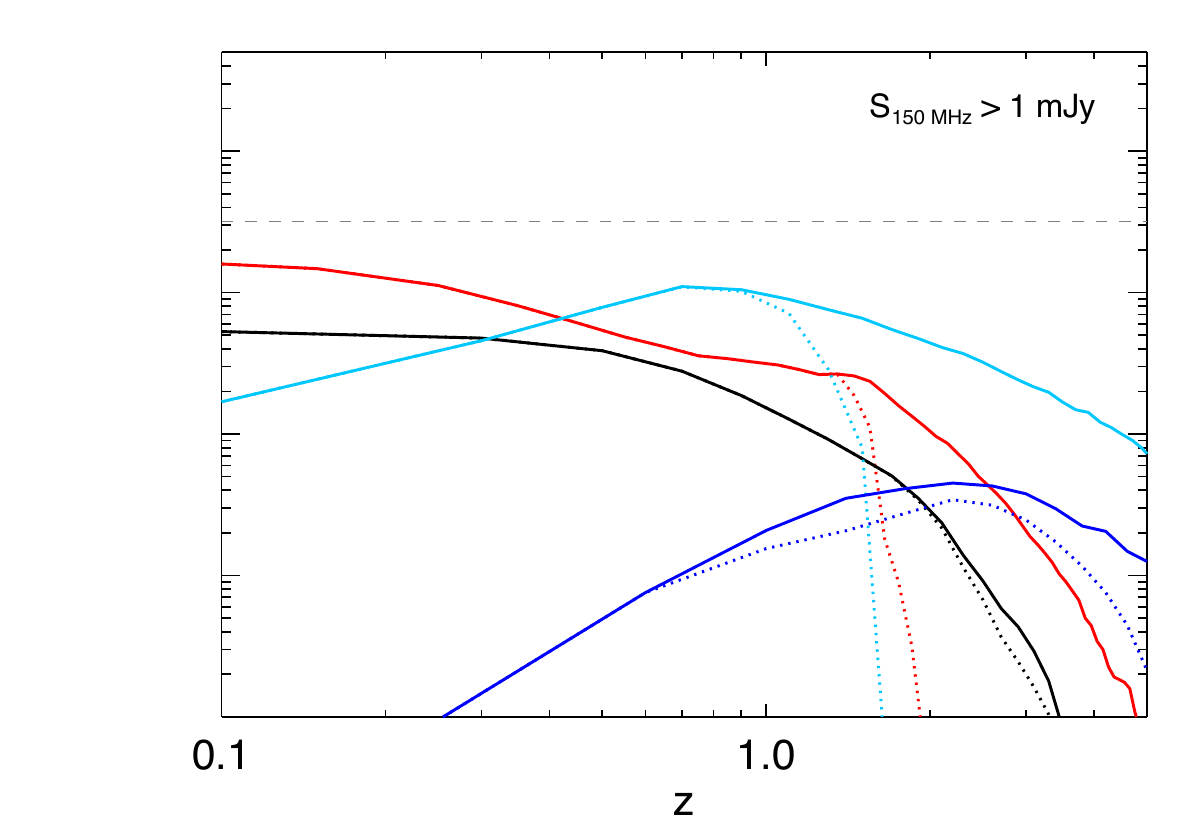} 
 \includegraphics[height=0.27\textwidth, trim=4.1cm 0cm 0.52cm 0cm, clip=false]{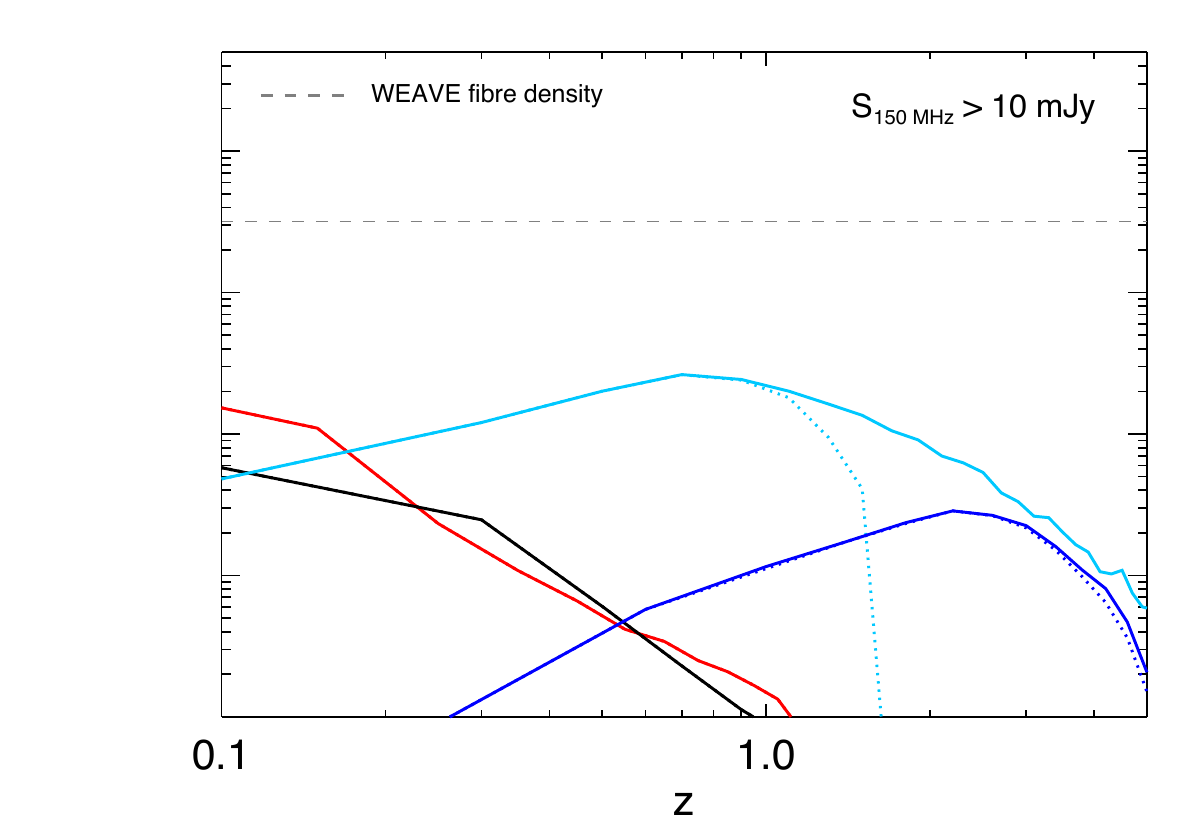}      
  \caption{The expected redshift distributions for each tier of WEAVE-LOFAR, including the Deep tier, at $S_{150} > 100$\,$\mu$Jy ({\bf Left}), the Mid tier ($S_{150} > 1$\,mJy; {\bf Centre}), and the Wide tier ($S_{150} > 10$\,mJy; {\bf Right}). Solid lines show the model redshift distribution while dotted lines of the same colour show the distribution of successful redshifts that we expect. The population has been divided to reveal the expected redshift distributions for each constituent of the source population according to the SKA Simulated Skies \citep{wilman08}, including star-forming galaxies (SFGs, in red), radio-quiet AGN (RQ-AGN, in black), high excitation radio galaxies (HEGs, in blue) and low-excitation radio galaxies (LEGs, light blue). 
   }
  \label{fig:dsmith_dndz}
\end{figure}

The redshift distribution of each sub-component of the faint radio source population, estimated using the SKA simulated skies \citep{wilman08} is shown for each tier of WEAVE-LOFAR in figure \ref{fig:dsmith_dndz}, from left to right in order of decreasing source density. The star forming galaxies are shown as red lines, the radio-quiet AGN are shown in black, while the high- and low-excitation radio galaxies are shown as blue and light-blue lines, respectively. Also overlaid are the redshift success rates that we expect to obtain in 1hr integrations for each population, indicated by the dotted lines of each colour. 

The redshift success rates have been estimated by taking the SKA simulated skies, alongside a simple but realistic model for each component of the faint radio source population, and accounting for a realistic sky spectrum, instrument throughput, fibre size, dust reddening and wavelength coverage, and assuming a default 1 hr integration on every source (though we plan to include deeper integrations of targets in the deep tier which don't yield a redshift in the first hour). The high redshift success rate in these simulations (which approaches 100 per cent at $z < 1$) is due to the fact that radio sources are predominantly emission line galaxies, or luminous ellipticals with strong 4000\,\AA\ breaks. The redshift success rate is also included in the right-hand panel of figure \ref{dsmith:lumz_fig}, which shows that the WEAVE-LOFAR sample will be predominantly star-forming galaxies at $z < 1.5$ and high redshift radio galaxies at $z > 2$.

\section{Science Highlights}
\label{sec:dsmith_cases}

The science case for WEAVE-LOFAR will be presented in more detail in Smith et al. (in prep), however in this section we present brief highlights from four of the main areas. 

\subsection{The star formation history of the Universe}

As figure \ref{fig:dsmith_sfr_sens} shows, the LOFAR surveys offer arguably the most effective way of tracing star formation in the Universe, since radio luminosity is roughly proportional to SFR \citep[e.g.][]{bell03}. Radio emission is unaffected by dust obscuration, and the high angular resolution available means that confusion effects do not limit the sensitivity \citep[unlike with Herschel, e.g.][]{oliver12}. While photometric, rather than spectroscopic, redshifts would be sufficient to trace the star formation density of the Universe, such redshifts are only reliable for objects with high signal-to-noise-ratio broadband detections and well-defined continuum breaks. Photometric redshifts are therefore inadequate for low-mass galaxies and/or those suffering from dust extinction, or indeed for those with bright emission lines, which comprise a large majority of the faint radio source population. The WEAVE spectra will also allow us to cleanly distinguish star forming galaxies from AGN, which can have similar flux densities at a range of redshift; spectra are therefore critical if we wish to unleash the immense power of LOFAR for studying star formation. 

WEAVE LOFAR is designed in such a way as to allow us to study the properties of star formation in galaxies as a function of stellar mass, environment, and redshift simultaneously, and with large statistical samples in every bin. This will enable us to study the influence of each factor on key properties of the galaxy population including the star formation rate density of the Universe \citep[e.g.][]{madau98,madau14}, the star formation rate function \citep[e.g.][]{smit12,cai14}, the fundamental metallicity relation \citep[e.g.][]{mannucci10,stott13}, and provide a new insight on the physics behind the far-infrared radio correlation \citep[e.g.][]{yun01,murphy09,murphy11,smith14}. Whilst we expect that individual spectra will detect only emission lines (i.e. not continuum), the uniformity of the WEAVE spectra, combined with the very large samples that we will observe, enables us to statistically detect the continuum properties of star-forming galaxies as well, reaching higher redshifts in better detail than has been previously possible. Those individual sources with continuum detections will also allow us to search for e.g. velocity offset absorption lines, thought to represent a `smoking gun'  of stellar\slash AGN feedback \citep[e.g.][]{chen10,davis12,geach14}, and it will also be possible to search for this signal statistically, averaging across populations \citep[e.g.][]{bradshaw13}.

\subsection{Accretion and AGN-driven feedback}

Since deep radio surveys can identify not only the classical double radio-loud sources, but also the much more abundant population of radio-quiet AGN at $z>1$ \citep{jarvis04,simpson06,seymour08,smolcic08}, they offer the opportunity to explore all aspects of AGN activity and evolution. By detecting radio-quiet AGN, independent of obscuration \citep[and knowing that most black hole growth is obscured; e.g.][]{martinez05}, radio surveys have the ability to make the most complete census of black hole accretion, including Compton-thick objects missed by X-ray surveys, allowing the poorly understood relationship between star formation and accretion to be studied further. WEAVE-LOFAR will provide new insight on the origin of the radio loud/quiet dichotomy \citep[e.g.][]{ivezic02,cirasuolo03,martinez11}. Furthermore, obtaining large and complete samples of AGN ensures that we are able to study the interplay between key properties of the AGN themselves, such as jet power \citep[probed by the radio luminosity; e.g.][]{hardcastle07,gurkan15}, or accretion rate \citep[as probed by emission line luminosity; e.g.][]{heckman04}, as well as their large-scale environments, and how these relationships vary with redshift. The complete redshift information for the radio-loud AGN population provided by WEAVE-LOFAR will be essential for determining the jet kinetic luminosity function, quantifying the amount of energy injected by jets of various powers into the intergalactic medium of their host galaxies.

Spectroscopy is essential if we wish to reliably discriminate between radiatively-efficient and -inefficient accretion \citep[the so-called ``quasar" and ``radio" modes;][]{croton06} through the presence or absence of high ionization emission lines \citep[see e.g.][]{osullivan15,best12}. Here again, spectroscopy provides vital input for testing galaxy formation models, which often assume that AGN feedback controls the evolution of galaxies. In our current understanding of the AGN feedback phenomenon, the quasar mode can drive gas out of the galaxy to terminate star-formation activity \citep{silk98}, while the radio (or ``jet") mode provides a lower rate of energy input that balances gas cooling, maintaining massive galaxies as ``old, red and dead" \citep{best06}. Recent studies have shown a decline in the space density of radio-mode AGN at $z<1$, and it has been suggested that this is related to the merger history of these galaxies \citep{rigby11,simpson12} and their gas supply \citep[e.g.][]{best14,williams15}. WEAVE spectroscopy is required to determine radio-selected AGN accretion mechanisms and provides essential data for testing models of AGN-galaxy co-evolution (e.g., the relationship between star formation and accretion over cosmic history). For example, tracking the evolution of radio-mode AGN and comparing it with the evolution of massive quiescent galaxies (i.e. their likely hosts) will be key for studying the quenching mechanism itself.

\subsection{The epoch of reionisation}

The recent discovery and subsequent study of a QSO at $z > 7$ have suggested that reionisation is incomplete 750\,Myr after the Big Bang \citep{mortlock11,bolton11}. However, the large cross-section for Lyman-$\alpha$ absorption means that the intergalactic medium becomes opaque for neutral fractions as low as $\sim$0.1 per cent, limiting the usefulness of optical studies. The 21\,cm hyperfine transition, by contrast, has a cross-section around $10^7$ times smaller, and LOFAR will be able to map the distribution of neutral gas at $z > 6$ provided sufficiently bright radio sources can be found at high redshift \citep{carilli02,mack12}. Current models predict one $S_{\mathrm{150 MHz}} > 10$\,mJy source at $z > 6$ every $\sim$200 deg$^2$, but these extremely valuable, ultra- rare sources are difficult to isolate from the overall population and will be more efficiently identified as part of a large-scale survey effort rather than through single-object follow-up, and LOFAR will be able to discover and identify them due to its large field of view and high survey speed. Assuming conservative $z > 6$ source densities (though recent work by Saxena et al. {\it in prep} suggests that the true numbers may be higher), a 10,000\,deg$^2$ survey would be expected to find $\sim50$ $z > 6$ radio galaxies (giving multiple lines of sight into this critical era of cosmic history), and would permit us to conduct 21\,cm absorption experiments. Larger samples are highly desirable, given that the brighter and higher-redshift sources are easier to probe for 21\,cm absorption, and that they provide more information about the reionisation process. A statistical sample of EoR sources will enable us to answer some of the most interesting questions about the reionisation process, such as ``how long did it take?", ``how clumpy was it?", and ``which sources were responsible?".

WEAVE-LOFAR will also be able to probe the demographics of the black hole population at these very large redshifts to answer important questions such as ``how do massive black holes and galaxies form?" and enabling us to determine the relationship between the black hole itself and the long-pursued massive and metal free stars (i.e. ``population III" stars). This approach is complementary to other searches for QSOs in the Epoch of Reionisation as the low-frequency radio surveys are also sensitive to the highly dust-obscured sources that may be missed by optical\slash near-IR or X-ray survey data.

\subsection{Radio galaxies and protoclusters}

Studying the formation of galaxy clusters and their member galaxies is key to our understanding of cosmology and galaxy evolution as they are ideal laboratories to test the standard model of structure and galaxy formation \citep{kravtsov12}. The most important epoch to study is the 3\,Gyr period around $z \approx 2$. During this period, the first cluster-sized halos collapsed \citep{chiang13}, forming the nuggets that will grow into massive clusters by the present day. Furthermore, cluster galaxies underwent a rapid period of star formation during this time, forming the bulk of the stars that now reside in massive cluster ellipticals \citep{delucia06}. The evolution of galaxies is affected by their surroundings, so by $z\approx 1$ there exists an intimate relationship between the properties of galaxies and their environments \citep[e.g.][]{dressler80,behroozi10,peng10,houghton15}. Observations of forming galaxies in this critical phase are therefore essential to obtain a complete understanding of how clusters and their member galaxies form.

This cluster formation process is still shrouded in mystery, despite a handful ($\sim15$) of protoclusters having been observed in detail. This is because current studies have not been able to place protoclusters in cosmological context: we do not know what their current halo mass is, nor how large each protocluster will grow. Without these key pieces of information it is hard to interpret what these galaxy overdensities are and how they fit into our theory of structure formation. The WEAVE-LOFAR IFU survey will transform our understanding of the early stages of cluster formation by measuring the already collapsed mass and estimating the present-day descendant cluster mass of 50 protoclusters at $1 < z < 2.5$. The protoclusters will be identified as groups of two or more radio galaxies within 1.5\,arc min from the WIDE tier of the WEAVE-LOFAR MOS survey, where we have the best sensitivity to the rarest objects. Both observational evidence \citep[e.g.][]{venemans07,wylezalek13,hatch14} and the latest simulations \citep[in particular those by][]{orsi16} agree that high redshift radio galaxies are beacons for protoclusters.

Numerical simulations show that the progenitors of massive galaxy clusters consist of many separate halos spread over ~25 Mpc at $z > 2$ \citep{chiang13,muldrew15,contini16}. The most massive of these halos could reach a few times $10^{13}$ to $10^{14}\,M_\odot$, and be signposted with a radio galaxy \citep{orsi16}. The rarest of the rare, ultra-massive protoclusters will contain several massive halos, each of which may be signposted with a radio-galaxy. Hence the association of several radio galaxies implies the presence of an agglomeration of several $>10^{13}\,M_\odot$ dark matter haloes that will eventually combine to form some of the most massive clusters in the Universe. The WEAVE-LOFAR IFU survey will not only locate them, but also put them into cosmological context for the first time. We will measure the collapsed dark matter halo mass of the radio galaxies by cross correlating the radio galaxies with the surrounding emission line galaxy population (either Lyman-$\alpha$ at $z > 2.1$ or [O{\sc ii}] at $z < 1.5$, taking advantage of WEAVE's broad wavelength coverage). The final descendant cluster mass can only be measured with a high-spectral resolution instrument, such as WEAVE. 

\section{Conclusions}
\label{sec:dsmith_concs}

We have described the WEAVE-LOFAR survey, which will use the thousand-fold multiplex and large contiguous wavelength coverage of the WEAVE instrument \citep{dalton12,dalton14}, scheduled for first light on the William Herschel Telescope in 2018, to produce more than $10^6$ spectra of sources selected at 150\,MHz from the LOFAR Surveys Key Science Project \citep{rottgering11}. WEAVE-LOFAR will have three tiers, in order to efficiently sample the luminosity-redshift plane, and to representatively probe the whole faint radio source population. WEAVE-LOFAR will be the primary source of spectroscopic redshifts and source classifications that are required to harness the vast star formation and accretion sensitivity of LOFAR, and will allow a very wide range of science topics to be addressed. These topics include the star formation history of the Universe, the interplay between different accretion modes, and their relationships with AGN-driven feedback, probing the epoch of reionisation, cosmology, and understanding cluster haloes and radio relics. The vast spectroscopic data set that will be produced will have immense legacy value, including e.g. allowing the necessary calibration of photometric redshifts for the {\it Euclid} mission and LSST. WEAVE-LOFAR will also include resolved spectroscopy of protocluster targets, essential for putting these growing overdensities into their cosmological context, and studying the influence of powerful AGN on their surrounding galaxies. Due to the demographics of the faint radio source population, we expect that observing low-frequency selected sources will have a high redshift success rate, approaching 100 per cent at $z < 1$. 

\begin{acknowledgements}
TS acknowledges support from the ERC Advanced Investigator programme NewClusters 321271. H.D. acknowledges financial support from the Spanish Ministry of Economy and Competitiveness (MINECO) under the 2014 Ramón y Cajal program MINECO RYC-2014-15686.

\end{acknowledgements}

\bibliographystyle{aa}  
\bibliography{dsmith_refs} 

\end{document}